\documentclass[reprint, amsmath,amssymb, aps]{revtex4-2}
\usepackage{graphicx}
\usepackage{physics}
\usepackage{dcolumn}
\usepackage{bm}
\usepackage{appendix}

\begin{document}

\preprint{APS/123-QED}

\title{Few-cycle excitation of atomic coherence: A closed-form analytical solution\\ 
beyond the rotating-wave approximation}

\author{Nazar Pyvovar}
\author{Bing Zeng}
\author{Lingze Duan}
 \email{nvp0004@uah.edu}
\affiliation{Department of Physics and Astronomy, The University of Alabama in Huntsville, Huntsville, AL 35899, USA.}

\date{\today}

\begin{abstract}
Developing an analytical theory for atomic coherence driven by ultrashort laster pulses has proved to be challenging due to the breakdown of the rotating wave approximation (RWA). In this paper, we present an approximate, closed-form solution to the Schrödinger equation that describes a two-level atom under the excitation of a far-off-resonance, few-cycle pulse of arbitrary shape without invoking the RWA. As an example of its applicability, an analytical solution for Gaussian pulses is explicitly given. Comparisons with numerical solutions validate the accuracy our solution within the scope of the approximation. Finally, we outline an alternative approach that can lead to a more accurate solution by capturing the nonlinear behaviors of the system.
\end{abstract}

\maketitle

\section{\label{sec:level1}Introduction}

Quantum coherent control (QCC) is of great importance in fundamental physics as well as a breadth of emerging applications \cite{Brif_2010}. With the advent of femtosecond and attosecond light sources, control of atomic coherence using ultrafast laser pulses with few optical cycles has attracted growing interest in recent years \cite{Genkin_1998, Jirauschek_2005, Xie_2009, Kumar_2012, Doslic_2006, Roudnev_2007, Wu_2007, Rostovtsev_2009, Jha_2010-1, Jha_2010-2, Jha_2010-3, Begzjav_2017, Arkhipov_2019, Zeng_2021}. Apart from its significance in quantum theories, ultrafast QCC has profound implications in practical applications. For example, in certain QCC schemes, using ultrashort, broadband pulses allows the first electronic states of molecules to be accessible and, at the same time, enables fast population transfer that occurs well within the typical collision times \cite{Sola_1999, Cai_2013}. Few-cycle pulses can also excite coherence on high-frequency transitions that enables efficient generation of extreme ultraviolet (XUV) radiations \cite{Scully_2008, Jha_2010-3}. 

Studying ultrafast QCC in the few-cycle regime faces unique challenges. The ultrashort pulse width invalidates the slowly-varying envelope approximation (SVEA) \cite{Leblond_2013} while the high peak field causes breakdown of the rotation-wave approximation (RWA) \cite{Begzjav_2017}. As a result, the well-established theoretical framework based on the optical Bloch equations and the area theorem ceases to apply \cite{Hughes_1998, Novitsky_2012}. Theoretical analysis has to rely on the Bloch equations or the Schrödinger equation in their original forms without simplifications, which are often highly nonlinear. This significantly increases the difficulty of developing analytical theories. In most cases, numerical simulations have to be used when dealing with few-cycle excitations \cite{Brown_1998, Kumar_2013, Novitsky_2012, Yang_2014}.

Meanwhile, there has been a continued effort to develop analytical theories for atomic coherence driven by few-cycle pulses \cite{Genkin_1998, Roudnev_2007, Wu_2007, Rostovtsev_2009, Jha_2010-1, Jha_2010-2, Jha_2010-3, Begzjav_2017, Arkhipov_2019, Zeng_2021}. A notably successful theory, proposed first by Rostovtsev \textit{et al.}, considers the coherence of a two-level atom under the excitation of a far-off-resonance strong ultrashort pulse \cite{Rostovtsev_2009}. Through a perturbative scheme, the model gives rise to a general solution of the Schrödinger equation without invoking the RWA. The original solution, however, is not in closed form, and the analysis of its features useful for practical applications still has to rely largely on numerical computations. Several attempts have been made to derive more explicit solutions under specific conditions \cite{Jha_2010-2,Jha_2014,Zeng_2021}. In particular, we have recently shown that a simple, closed-form analytical solution of the Schrödinger equation can be obtained for few-cycle \textit{square} pulses \cite{Zeng_2021}.

In this paper, we expand our model to include arbitrary pulse shapes while retaining an explicit, closed-form solution. Excitation by a few-cycle Gaussian pulse is analyzed as an example of the general solution. The accuracy of our solution is verified by comparing it to the exact numerical solution of the general Schrödinger equation. Finally, an alternative approach is suggested to simplify the theory, which can potentially lead to a more accurate solution with a closer representation of the nonlinear behaviors of the system and a broader scope of applicability. 

\section{Closed-Form Solution}
\subsection{General Model}

Our general model follows the theoretical framework described in Ref. \cite{Rostovtsev_2009}. A quick outline is given below. We consider a two-level system (TLS) under the influence of an electromagnetic field. The Hamiltonian of the system is
\begin{equation}
    \hat{H}=\hbar\omega_c\ket{c}\bra{c}-\mu \mathcal{E}(t)\ket{c}\bra{d}-[\mu\mathcal{E}(t)]^*\ket{d}\bra{c},
\end{equation}
where $\ket{c}, \ket{d}$ are upper and lower levels, respectively, $\omega_c$ is the transition frequency, $\mathcal{E}(t)$ is the electric field, and $\mu$ is the dipole moment of the system. We are interested in the electric field of the form $\mathcal{E}(t)=\Omega(t)\cos(\omega t+\phi)$, where $\Omega(t)$ is the pulse envelope function and $\phi$ is carrier-envelope phase (CEP).

With this Hamiltonian, the equations of motion for the system are given by
\begin{subequations}
\begin{align}
&\dot C(t)=-i\Omega(t)\cos(\omega t+\phi)e^{i\omega_c t}D(t)\label{Cdot1},\\
&\dot D(t)=-i\Omega^*(t)\cos(\omega t+\phi)e^{-i\omega_c t}C(t)\label{Ddot1},
\end{align}
\label{mainsystem}
\end{subequations}
where $C(t)$ and $D(t)$ are the probabilities of $\ket{c}$ and $\ket{d}$, respectively, i.e., $\ket{\Psi}=C(t)e^{-i\omega_ct}\ket{c}+D(t)\ket{d}$.

It proves useful to introduce the following quantity to simplify our equations at this point
\begin{equation}
    \theta(t)=\int_{-\infty}^t \Omega(t')\cos(\omega t'+\phi)e^{i\omega_c t'}dt'\label{theta}.
\end{equation}
With this definition, the equations \eqref{Cdot1}, \eqref{Ddot1} become
\begin{subequations}
\begin{align}
&\dot C(t)=-i\dot\theta(t) D(t)\label{Cdot2},\\
&\dot D(t)=-i\dot\theta^*(t) C(t)\label{Ddot2}.
\end{align}
\label{system}
\end{subequations}
In this paper, we are only interested in the case when electromagnetic field exciting the system is an ultrashort pulse that is non-zero within a finite time interval. For this reason, all quantities hereafter are considered only in a finite time interval $t\in [0, \tau]$. In particular, $\Omega(t)$ is only non-zero inside of this interval, and hence all the dynamics of the system occur exclusively within the interval as well. With this in mind, we can replace the lower integration limit in \eqref{theta} with $0$.


By introducing the quantity $f=C/D$, the Schrödinger equation \eqref{system} can be simplified to  
\begin{equation}
    \dot f(t)=i\dot\theta^*(t) f^2(t)-i\dot\theta \label{fnonlinear}.
\end{equation}
The zeroth-order solution of \eqref{fnonlinear} is obtained by neglecting the $f^2(t)$ term \cite{Rostovtsev_2009}, which yields
\begin{equation}
    f_0(t)=-i\theta(t).
    \label{f0}
\end{equation}
Now, define the first-order approximation $f_1(t)$ as the solution of \eqref{fnonlinear} that satisfies $(f_1(t)-f_0(t))^2\ll 1$. It is then straightforward to show that $f_1(t)$ satisfies the equation
\begin{equation}
    \dot f_1(t)=2\theta(t)\dot\theta^*(t)f_1(t)+i\theta^2(t)\dot\theta^*(t)-i\dot\theta(t)\label{f1}.
\end{equation}
In this way, we can define a sequence of approximations to \eqref{fnonlinear}. Specifically, if the $k$th-order approximation $f_k$ is found, then the $(k+1)$th-order approximation $f_{k+1}$ is the solution of $f$ that meets the condition $(f_{k+1}-f_k)^2\ll 1$, which satisfies the equation
\begin{equation}
    \dot f_{k+1}=2if_k(t)\dot\theta^*(t)f_{k+1}(t)-if_k^2(t)\dot\theta^*(t)-i\dot\theta(t)\label{fk}.
\end{equation}
For far off-resonance, however, it has been shown that the first approximation \eqref{f1} is already quite accurate. Therefore, we focus on \eqref{f1} in the current work. The general solution to \eqref{f1} has been found to be \cite{Zeng_2021}
\begin{equation}
f_1(t)=-\frac{i}{2}[\theta(t)+\int_{0}^{t}\dot\theta(t')e^{\alpha(t', t)}dt'],
\label{f1solution}
\end{equation}
where $\alpha(t', t)=2\int_{t'}^{t}\theta(t'')\dot\theta^{*}(t'')dt''$. 

\subsection{Approximate Solution}
Although the solution \eqref{f1solution} is generic, the integral on the right-hand side cannot be evaluated in closed-form for any realistic pulse shape.

To find a sufficiently accurate closed-form solution of \eqref{f1}, we can expand $\alpha(t', t)$ in powers of $\delta=t-t'$ and approximate
\begin{equation}
\alpha(t', t)=2\int_{t'}^{\delta+t'}\theta\dot\theta^*dt''=2\theta(t)\dot\theta^{*}(t)\cdot [t-t'],
\end{equation}
to the leading term in $\delta$.

Using the fact that all quantities considered here are assumed to be limited to a finite time interval $t\in[0, \tau]$, we can further approximate the leading term above as
\begin{equation}\alpha(t', t)=2(t-t')\frac{1}{\tau}\int_0^\tau\theta(t'')\dot\theta^*(t'')dt''=-i\alpha_0(t-t'),\end{equation}
where
\begin{equation}
\alpha_0=\frac{2i}{\tau}\int_0^\tau\theta(t'')\dot\theta^*(t'')dt''
\label{alpha0}
\end{equation}
is a constant (the average of $\alpha(t', t)$), and $\tau$ is once again the width of the few-cycle driving pulse. This allows us to simplify \eqref{f1solution} by noting
\begin{align}
&\int_{-\infty}^{t}\dot\theta(t')e^{\alpha(t',t)}dt'\nonumber\\
&=\int_{-\infty}^t\Omega(t')\cos(\omega t'+\phi)e^{i\omega_ct'}
e^{-i\alpha_0(t-t')}dt'\nonumber\\
&=e^{-i\alpha_0t}\int_{-\infty}^t\Omega(t')\cos(\omega t'+\phi)e^{i(\omega_c+\alpha_0)t'}dt'.
\label{12}
\end{align}

At this point, it is convenient to re-define $\theta$ to be a function of $\omega_c$ as well as $t$ so that
\begin{equation}
\theta(\omega_c, t)=\int_{-\infty}^{t} \Omega(t')\cos(\omega t'+\phi)e^{i\omega_{c}t'}dt'.
\label{theta2}
\end{equation}
With this notation, we can write \eqref{12} simply as
\begin{equation}
\int_{-\infty}^{t}\dot\theta(t')e^{\alpha(t', t)}dt'=e^{-i\alpha_0 t}\theta(\omega_c+\alpha_0, t).
\label{thetatrick}
\end{equation}

Substituting \eqref{thetatrick} back into \eqref{f1solution}, we have a closed-form solution for pulses of arbitrary shapes:
\begin{equation}
    \Tilde f_1(t)=-\frac{i}{2}[\theta(\omega_c, t)+e^{-i\alpha_0 t}\theta(\omega_c+\alpha_0, t)],
    \label{f1closed}
\end{equation}
where $\alpha_0$ is a constant given by \eqref{alpha0}. From now on, tilde will be used to denote the approximate solutions, while a notation without tilde denotes an exact solution.

\subsection{Sequence of Approximate Solutions}

In order to obtain a more accurate solution of $\eqref{fnonlinear}$, we use the recursive relation \eqref{fk} to get a sequence of closed-form solutions $\Tilde{f}_k$ similar to \eqref{f1closed}. Taking the limit of $k\to \infty$ will lead us to an improved solution.

First, note that the general solution to the linear equation \eqref{fk} is:
\begin{align}
    f_{k+1}&=e^{2i\int_{-\infty}^t f_k(t')\dot\theta^*(t')dt'}\times\\ \nonumber
    &\left(\int_{0}^t e^{-2i\int_{-\infty}^{t'} f_k(t'')\dot\theta^*(t'')dt''}[-if_k^2(t')\dot\theta^*(t')-i\dot\theta(t')]dt'\right),
\end{align}
where the initial condition $f(0)=0$ has been used. Integrating the first integrand by parts and bringing the outside exponential inside the integral yield
\begin{align}
    f_{k+1}(t)=\frac{1}{2}\left(f_k(t)-\int_{0}^t e^{\beta_k(t', t)}\left(\dot f_k(t')+2i\dot\theta(t')\right)dt'\right),
    \label{fk+1}
\end{align}
where
\begin{align}
    \beta_k(t', t)=e^{2i\int_{t'}^t f_k(t')\dot\theta^*(t')dt'}.
    \label{betak}
\end{align}

Using the same idea as with \eqref{f1closed}, we approximate $\beta_k$ with a linear function of $(t-t')$
\begin{equation}
    \beta_k(t', t)=-i\alpha_k(t-t'),\;\; \alpha_k=-\frac{2}{\tau}\int_0^\tau\dot\theta^*(t')\Tilde f_k(t')dt'.
\end{equation}
Notice that we have used $\Tilde f_k$, which is yet to be found, in the expression for $\alpha_k$ above. Substituting this back into \eqref{fk+1}, and using the same trick as in \eqref{thetatrick}, we have
\begin{align}
    \Tilde f_{k+1}(t)=\frac{1}{2}\{\Tilde f_k(t)&-2ie^{-i\alpha_k t}\theta(\omega_c+\alpha_k, t)\nonumber\\
    &-\int_{0}^t e^{-i\alpha_k(t-t')}\dot{\Tilde{f}}_k(t')dt'\},
    \label{recursion}
\end{align}
where we again note that all $f_k$ are tilded.

From \eqref{recursion}, it is possible to derive an accurate solution for $\Tilde{f}_{k+1}$ and then to take the limit $k\to\infty$. Although this would be the cleanest way to derive the limiting solution, it is very tedious. Here, we offer a more elegant approach. Assume that all functions in \eqref{recursion} are continuous and that taking the limit commutes with integration. Take the limit $k\to\infty$ on both sides and denote $\Tilde{f}_{\infty}=\displaystyle{\lim_{k\to\infty}}\Tilde{f}_k$ and $\alpha_\infty=\displaystyle{\lim_{k\to\infty}}\alpha_k$. We then have
\begin{align}
    \frac{1}{2}\Tilde{f}_\infty(t)=&-ie^{-i\alpha_\infty t}\theta(\omega_c+\alpha_\infty, t)\nonumber\\
    &-\frac{1}{2}e^{-i\alpha_\infty t}\int_{0}^t e^{i\alpha_\infty t'}\dot{\Tilde{f}}_\infty(t')dt'.
\end{align}

Introducing $g(t)=e^{i\alpha_\infty t}\Tilde{f}_\infty(t)+i \theta(\omega_c+\alpha_\infty, t)$ and integrating the exponential by parts lead to
\begin{equation}
    \frac{1}{2}g(t)=-\frac{1}{2}\int_0^t
    \left(\dot g(t')-i\alpha_\infty (g(t')-i\theta(\omega_c+\alpha_\infty, t'))\right)dt',
\end{equation}
and, after differentiation,
\begin{equation}
    \dot g(t)=\frac{i\alpha_\infty}{2}g(t)+\frac{\alpha_\infty}{2}\theta(\omega_c+\alpha_\infty, t),
\end{equation}
which yields
\begin{equation}
    g(t)=e^{\frac{i\alpha_\infty}{2}t}\left(\int_0^t e^{\frac{-i\alpha_\infty}{2}t'}\frac{\alpha_\infty}{2}\theta(\omega_c+\alpha_\infty, t')dt'+C\right).
\end{equation}
Applying the initial condition $g(0)=0$ and integrating by parts, $g(t)$ is simplified to
\begin{equation}
    g(t)=i\theta(\omega_c+\alpha_\infty, t)-ie^{\frac{i\alpha_\infty}{2}t}\theta(\omega_c+\frac{\alpha_\infty}{2}, t).
\end{equation}
Substituting $g(t)$ back to $\Tilde{f}_\infty$, we have
\begin{equation}
    \Tilde{f}_\infty(t)=-i e^{-i\lambda t}\theta(\omega_c+\lambda, t),
\end{equation}
where $\lambda=\alpha_\infty/2$ is a constant that we will find next.

The downfall of this shortcut derivation is that we do not immediately have an expression for $\lambda$ (or $\alpha_\infty$). To find $\lambda$, let us consider $\lambda$ as an arbitrary constant and try to identify the value of $\lambda$ that would make \eqref{finfinity} most accurate. Substituting \eqref{finfinity} back into \eqref{fnonlinear}, we find
\begin{equation}
    \lambda=i\dot\theta^*(\omega_c, t)e^{-i\lambda t}\theta(\omega_c+\lambda, t),
\end{equation}
which provides an equation the optimal $\lambda$ should satisfy. Notice that the right hand side is time-dependent, so averaging it and using $\dot\theta^*(\omega_c, t)e^{i\nu t}=\dot\theta^*(\omega_c+\nu, t)$ lead us to the equation for the optimal $\lambda$,
\begin{equation}
   \lambda=\frac{i}{\tau}\int_0^\tau \dot\theta^*(\omega_c+\lambda, t')\theta(\omega_c+\lambda, t')dt'.
    \label{alphainfinity}
\end{equation}

Admittedly, \eqref{alphainfinity} does not provide a closed-form expression for $\lambda$. However, the role of $\lambda$ is as a particularly suitable constant that makes the approximate solution of the form \eqref{finfinity} most accurate. The fact that the form \eqref{finfinity} is closed-form is the crucial point here.

Thus, we can summarize our solution for the Schrödinger equation as follows:
\begin{equation}
    \Tilde{f}_\infty(t)=-i e^{-i\lambda t}\theta(\omega_c+\lambda, t),
    \label{finfinity}
\end{equation}
where
\begin{equation}
\lambda=\frac{i}{\tau}\int_0^\tau \dot\theta^*(\omega_c+\lambda, t')\theta(\omega_c+\lambda, t')dt',
\end{equation}
and
\begin{equation}
\theta(\nu, t)=\int_{-\infty}^{t} \Omega(t')\cos(\omega t'+\phi)e^{i\nu t'}dt'.
\end{equation}

\section{Gaussian Pulse Excitation}

\begin{figure*}[t]
\includegraphics[width=0.45\linewidth]{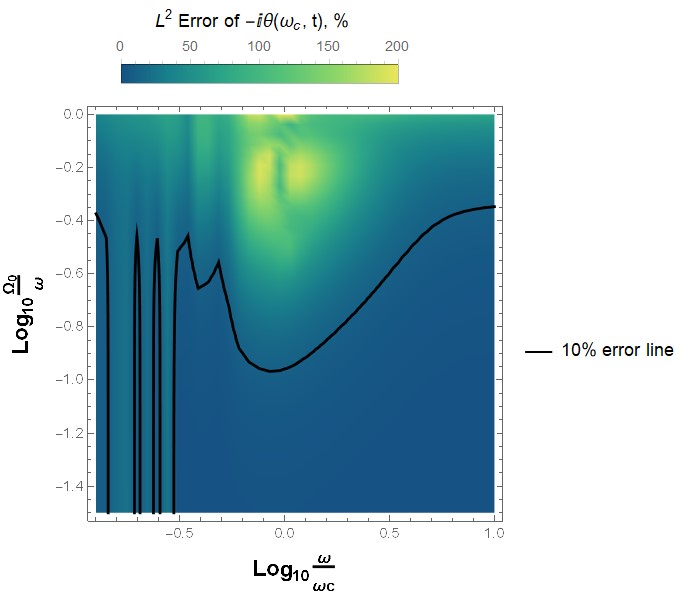}
\includegraphics[width=0.45\linewidth]{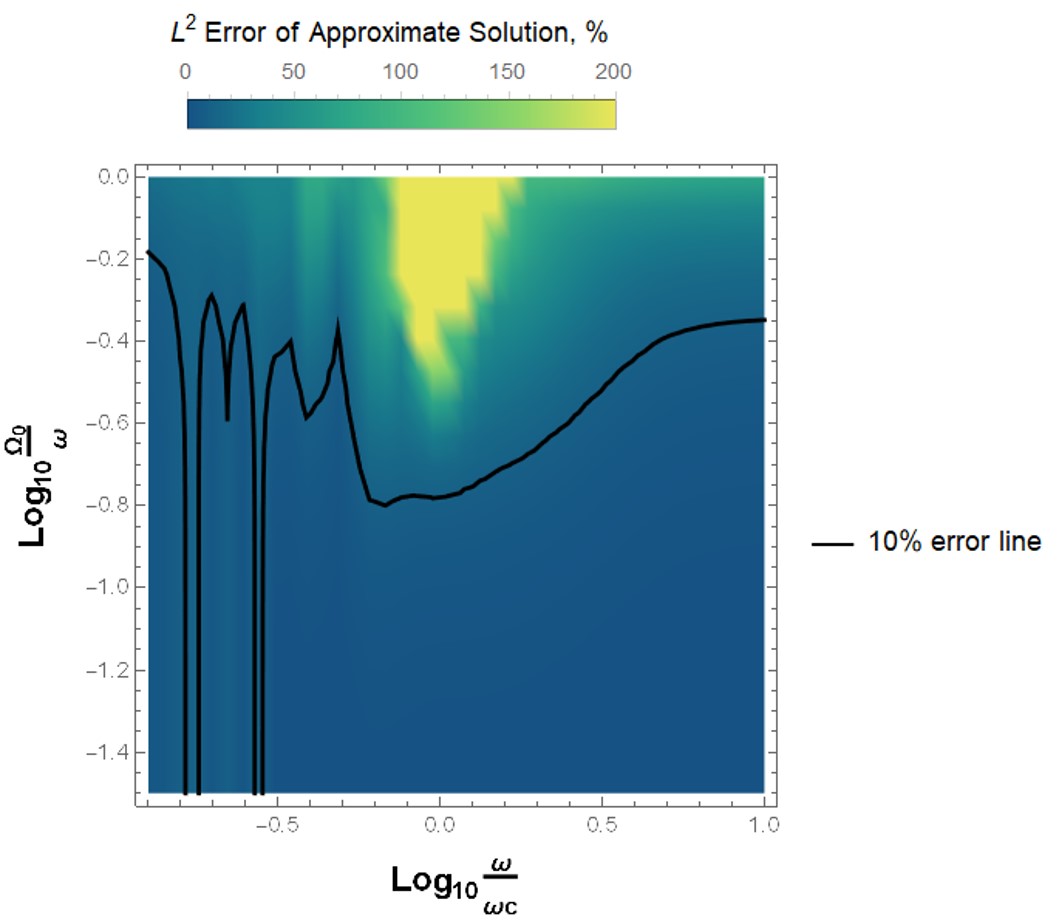}
\includegraphics[width=0.45\linewidth]{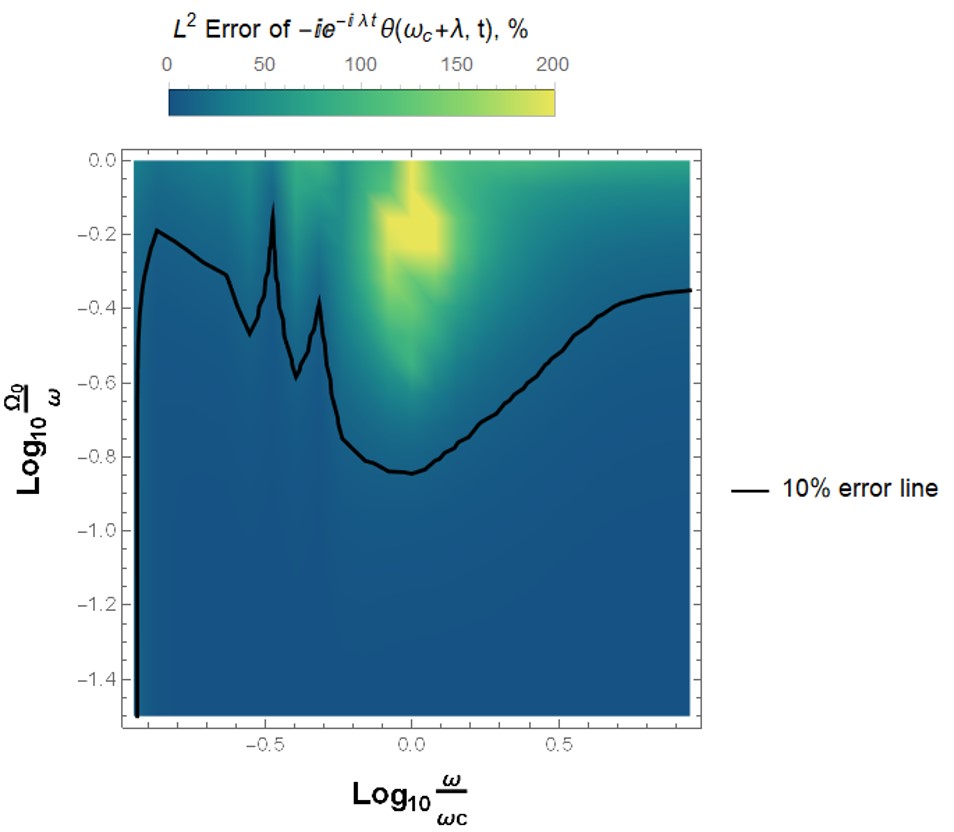}
\includegraphics[width=0.45\linewidth]{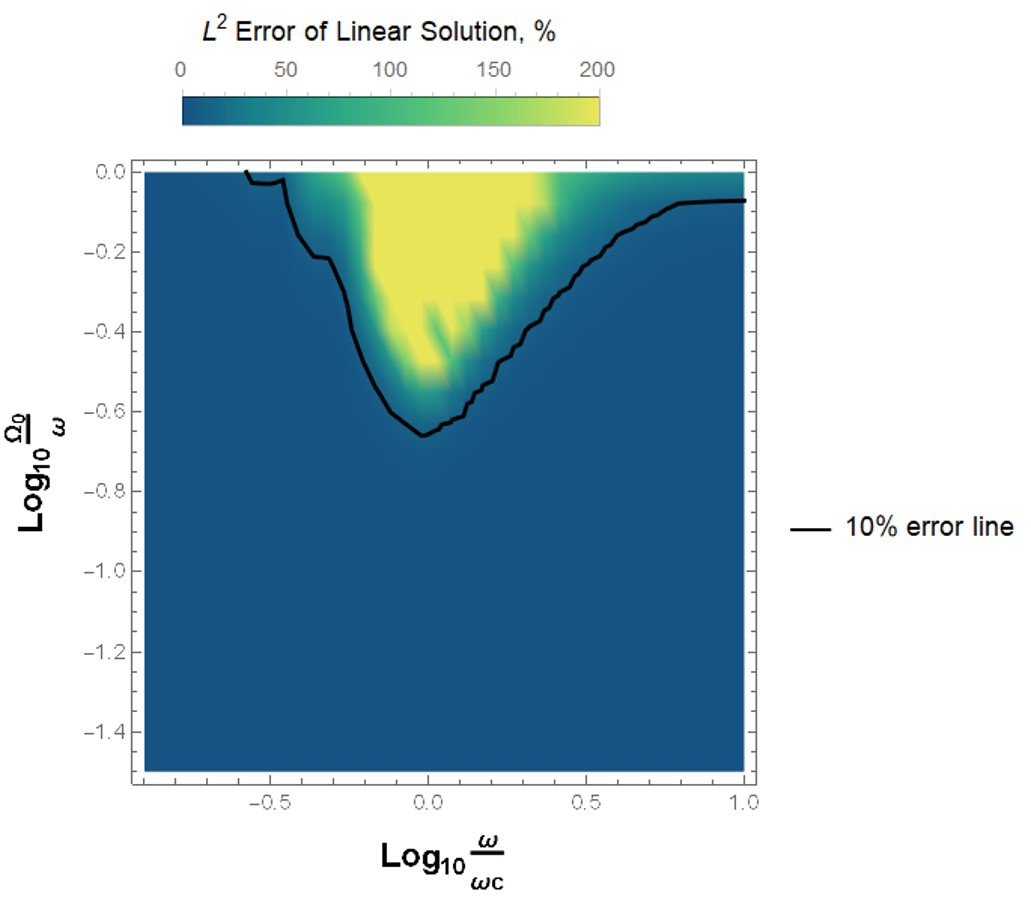}
\caption{The percent $L^2$ error of the respective approximate solution as compared to the exact numerical solution of \eqref{fnonlinear}. Top left: the zero-order approximation \eqref{f0}. Top right: the approximate solution \eqref{f1closed}. Bottom left: the limiting approximate solution \eqref{finfinity}. Bottom right: the exact solution of the linear equation \eqref{f1}. The solid black line is the contour on which the error is equal to 10\%.}
\end{figure*}
Let us study the conditions for which our solutions \eqref{f1closed} and \eqref{finfinity} are applicable, as well as their accuracy.

First of all, it should be pointed out that the main result of Ref. \cite{Zeng_2021} can be obtained from our solution \eqref{f1closed}. For that, let us assume $\alpha_0\ll \omega_c$ so that $\theta(\omega_c+\alpha_0, t)=\theta(\omega_c, t)$. In this case, denoting $\alpha_0=\eta^2\omega_c$, equation \eqref{f1closed} becomes
\begin{equation}
\Tilde{f}_1(t)=-\frac{i}{2}\left(1+e^{-i\eta^2\omega_ct}\right)\theta(\omega_c, t),
\end{equation}
which is exactly the equation (16) in Ref. \cite{Zeng_2021}. The latter solution was shown to be accurate for a square pulse under the condition \cite{Zeng_2021} 
\begin{equation}
    \left(\frac{\omega}{\omega_c}\right)^2+\left(\frac{\Omega_0}{\omega}\right)^2\ll 1
    \label{applicability}.
\end{equation}

Next, we examine how accurate the solution \eqref{f1closed} and its generalization \eqref{finfinity} are for more realistic pulses. As a quantity that expresses the accuracy of a solution, we choose the $L^2$-norm of the deviation of the solution from the exact solution. More specifically, the ratio of the latter quantity to the $L^2$-norm of the exact solution itself - we call this ratio the relative $L^2$ error. We numerically calculate this quantity for a Gaussian envelope centered in the middle of the pulse 
\begin{equation}
\Omega(t)=\Omega_0\exp{-\frac{(t-\frac{\tau}{2})^2}{2\sigma^2}},
\end{equation}
where we mainly explore the dependence of the relative error described above on the quantities $\Omega_0/\omega$ and $\omega_c/\omega$, since they were shown in Ref. \cite{Zeng_2021} to be the relevant parameters that determine applicability of a solution.

In Fig. 1, we plot the relative error described above for the approximate solutions \eqref{f0}, \eqref{f1closed} and \eqref{finfinity}. As one can see, our most recent solution is the most accurate of the three, especially in the region where the previous solutions did not apply - for the high $\omega_c/\omega$ ratios. At the same time, we can see that the applicability conditions are generally the same as \eqref{applicability}. Namely, the system must be far off resonance.

However, a slight modification of \eqref{applicability} is in order. Notice that our solution applies not only for the situations when $\omega/\omega_c\ll 1$, but also $\omega_c/\omega\gg 1$. As a result, the following condition is more appropriate
\begin{equation}
    \min\left(\left(\frac{\omega_c}{\omega}\right)^2, \left(\frac{\omega}{\omega_c}\right)^2\right)+\min\left(\left(\frac{\Omega_0}{\omega}\right)^2, \left(\frac{\omega}{\Omega_0}\right)^2\right)\ll 1.
    \label{newapplicability}
\end{equation}

As demonstrated in Fig. 1, our solution accurately predicts the behavior of the TLS driven by a realistic Gaussian pulse. While the solution applies to arbitrary pulse shapes, a Gaussian pulse offers an exclusively convenient property under the current context, namely, the function $\theta(\omega_c, t)$ defined by \eqref{theta2} can be evaluated explicitly in closed-form for a Gaussian pulse,

\begin{widetext}
\begin{equation}
\theta_{\text{Gaussian}}(\omega_c, t)=\frac{\sigma\Omega_0}{4}\left(\zeta(t, \omega)e^{i\phi}+\zeta(t, -\omega)e^{-i\phi}\right),
\label{thetaGauss}
\end{equation}
\begin{equation}
    \zeta(t, \omega)=e^{i\frac{\nu+\omega}{2}-\frac{1}{2}\sigma^2(\nu+\omega)^2}\left(\text{erf}\left(\frac{1+2i\nu\sigma^2+2i\sigma^2\omega}{2\sqrt{2}\sigma}\right)-\text{erf}\left(\frac{1-2t+2i\nu\sigma^2+2i\sigma^2\omega}{2\sqrt{2}\sigma}\right)\right).
\end{equation}
\end{widetext}

Such closed-form solutions have not been found for other common pulse shapes, such as the Lorentzian or the hyperbolic secant pulses.

\section{Discussion}

Another insightful comparison of \eqref{finfinity} is with the classical rotating-wave approximation(RWA) regime. Under the RWA, the equations of motion \eqref{mainsystem} become
\begin{align}
    \dot C(t)=-\frac{i}{2}\Omega(t)D(t),\\
    \dot D(t)=-\frac{i}{2}\Omega(t)C(t).
\end{align}
Introducing the area of the pulse $A(t)=\int_0^t\Omega(t')dt'$, we can find a general solution of the TLS in the form of
\begin{align}
    C(t)=-i\sin\frac{A(t)}{2},\\
    D(t)=\cos\frac{A(t)}{2}.
\end{align}
Since $f(t)=C(t)/D(t)$, we have 
\begin{equation}
f=-i\tan\frac{A}{2}
\label{RWAf}
\end{equation}
as the general solution under the RWA.

To appreciate the similarity of our solution, notice that we have the closed-form solution \eqref{finfinity} and the relation
\begin{equation}
\theta(\omega_c, t)=\frac{e^{i\phi}\Tilde{A}(\omega_c+\omega, t)+e^{-i\phi}\Tilde{A}(\omega_c-\omega, t)}{2}.
\end{equation}
Under RWA, $\Tilde{A}(\omega_c+\omega, t)=0$ and $\Tilde{A}(\omega_c-\omega, t)=A(t)$, which means that up to a phase factor, our solution gives
\begin{equation}
\Tilde{f}_\infty(t)=-i \frac{A}{2}.
\label{RWA-our}
\end{equation}
This suggests that, near resonance (which is outside of its region of applicability), our solution corresponds to the linearization of the RWA solution. This apparently is the result of multiple linearizations we have taken in deriving $\Tilde{f}_k$ to obtain the closed-form solution. With this observation in mind, we aim to overcome the oversimplification of our approximations and try to capture more of the nonlinear behaviors of the original Schrödinger equation. One approach is outlined as follows. 

Consider again the general equation \eqref{fnonlinear}
and use $\dot\theta(t)=\Omega(t)\cos(\omega t+\phi)e^{i\omega_c t}$. Taking $g(t)=\Omega(t)\cos(\omega t+\phi)$ and multiplying through by $e^{-i\omega_c t}$, we arrive at
\begin{equation}
    e^{-i\omega_c t}\dot f(t)=ig(t)e^{-2i\omega_c t}f^2(t)-ig(t).
\end{equation}
Substituting $z(t)=e^{-i\omega_c t}f(t)$ into the above equation leads to
\begin{equation}
    \dot z(t)=ig(t)z^2(t)-i\omega_c z(t)-ig(t).
    \label{zequation}
\end{equation}
Now expanding $z(t)$ as a perturbation series in $\omega_c$, we obtain for the zeroth order
\begin{equation}
\dot z_0(t)=ig(t)(z_0^2(t)-1),
\end{equation}
which gives
\begin{equation}
z_0(t)=\cosh^{-1}\left(i\int_0^t g(t)dt\right).
\label{z0}
\end{equation}
Meanwhile, the $n$th-order can be written as
\begin{equation}
    \dot z_{n}(t)=2ig(t)z_{n}(t)z_0(t)+i\displaystyle{\sum_{j=1}^{n-1}}z_j(t)z_{n-j}(t)-iz_{n-1}(t).
\end{equation}

The general solution to this linear equation, given the initial condition $z(0)=0$, is:
\begin{equation}
    z_{n}(t)=ie^{w(t)}\int_0^te^{-w(t')}\left(\displaystyle{\sum_{j=1}^{n-1}}z_j(t')z_{n-j}(t')-z_{n-1}(t')\right)dt',
    \label{zn}
\end{equation}
where $w(t)=2i\int_0^tg(t')z_0(t')dt'$. Together with \eqref{z0}, \eqref{zn} defines a sequence of functions that converge to the solution, similar to the sequence $f_k(t)$ discussed in Section II.

Note that, compared to the linearized sequence $f_k(t)$, the sequence of $z_n(t)$ is a much closer approximation to the accurate solution. Even the first-order approximation \eqref{z0} captures the nonlinear behavior of the system. If a sequence of closed-form approximations to \eqref{zn} could be found, similar to the sequence $\Tilde{f}_k$ presented in Section II, the limit of such sequence $z_{\infty}$ would be a promising approximation for capturing the nonlinear behaviors of the system. It seems plausible that a combination of such solution with the solution \eqref{finfinity} would allow us to extend the conditions of applicability of \eqref{finfinity} to the case when $\omega\sim\omega_c,\; \omega\sim\Omega_0$, and capture RWA better than a simple linear approximation. However, so far, we have been unable to find any approach to obtain a closed-form solution in this way. The difficulty is precisely the nonlinear features embedded into each \eqref{zn}.

\section{Conclusion}

In this work, we have developed a closed-form analytical solution of the Schrödinger equation that describes a TLS driven by a far-off-resonance few-cycle pulse without using the RWA. Unlike our previous solution, which only holds for square pulses \cite{Zeng_2021}, the current solution is generalized to arbitrary pulse shapes. We have identified the conditions under which our solution accurately predicts the behaviors of the system, and have demonstrated that the new solution offers improved accuracy compared to the previous solution under similar conditions. We have also applied the general solution to Gaussian pulses as a demonstration of its applicability. Finally, we suggest a possible alternative approach that can lead to a more accurate solution by capturing the nonlinear behaviors of the system and extending the applicability of the solution to the case of $\omega\sim\omega_c,\; \omega\sim \Omega_0$. This work lays out a potential pathway toward an analytical theory for ultrafast QCC beyond the RWA.

\nocite{*}
\bibliography{apsbib}

\end{document}